# Stable Tree Labelling for Accelerating Distance Queries on Dynamic Road Networks

Henning Koehler
Massey University
Palmerston North, New Zealand
H.Koehler@massey.ac.nz

Muhammad Farhan
Australian National University
Canberra, Australia
muhammad.farhan@anu.edu.au

Qing Wang
Australian National University
Canberra, Australia
qing.wang@anu.edu.au

## ABSTRACT

Finding the shortest-path distance between two arbitrary vertices is an important problem in road networks. Due to real-time traffic conditions, road networks undergo dynamic changes all the time. Current state-of-the-art methods incrementally maintain a distance labelling based on a hierarchy among vertices to support efficient distance computation. However, their labelling sizes are often large and cannot be efficiently maintained. To combat these issues, we present a simple yet efficient labelling method, namely *Stable Tree Labelling* (STL), for answering distance queries on dynamic road networks. We observe that the properties of an underlying hierarchy play an important role in improving and balancing query and update performance. Thus, we introduce the notion of *stable tree hierarchy* which lays the ground for developing efficient maintenance algorithms on dynamic road networks. Based on stable tree hierarchy, STL can be efficiently constructed as a 2-hop labelling. A crucial ingredient of STL is to only store distances within subgraphs in labels, rather than distances in the entire graph, which restricts the labels affected by dynamic changes. We further develop two efficient maintenance algorithms upon STL: *Label Search algorithm* and *Pareto Search algorithm*. Label Search algorithm identifies affected ancestors in a stable tree hierarchy and performs efficient searches to update labels from those ancestors. Pareto Search algorithm explores the interaction between search spaces of different ancestors, and combines searches from multiple ancestors into only two searches for each update, eliminating duplicate graph traversals. The experiments show that our algorithms significantly outperform state-of-the-art dynamic methods in maintaining the labelling and query processing, while requiring an order of magnitude less space.

## 1 INTRODUCTION

Road networks are dynamic, typically modeled as a weighted dynamic graph $G = (V, E, \phi)$, where vertices $V$ represent intersections, edges $E$ represent roads between intersections, and edge weights $\phi$ represent information that may evolve over time due to changing traffic conditions, e.g., travel time. Given two arbitrary vertices $u, v \in V$, computing their shortest-path distance, i.e., *distance query*, is arguably one of the most widely performed tasks in real-world applications, such as helping drivers' or autonomous cars to find a shortest-path, matching taxi drivers with passengers, optimizing delivery routes with multiple pick-up and drop-off points that change dynamically, or providing recommendation on $k$-nearest POIs to their customers [7, 11, 15, 20, 28]. For example, ride-hailing companies like Uber and Lyft need to compute millions of shortest-path distances to optimize routes for drivers under dynamic traffic conditions. This helps minimize wait times and ensure efficient pick-up and drop-off services, especially when traffic patterns change due to congestion or road closures. By frequently updating shortest-path distances, these companies ensure that drivers follow the most efficient routes to reach passengers quickly and provide optimal service [17, 29].

In static road networks, a plethora of approaches have been developed for answering distance queries, A classical approach is to run the uni- or bi-directional Dijkstra's algorithms [23, 26]. However, these methods can take several seconds to answer a single query on large road networks, which is impractical for real-world, time-sensitive applications where speed is crucial. To accelerate query response times, numerous methods have been developed [1, 2, 4, 6, 8, 9, 12, 13, 15, 16, 18, 19, 21, 24, 25, 33], which can be broadly classified into two categories: 1) *search-based methods* [6, 8, 13, 15, 19, 24, 25, 33], and 2) *labelling-based methods* [1, 2, 4, 9, 12, 18, 21]. Among search-based methods, *Contraction Hierarchy* (CH) [13] has demonstrated outstanding performance in practice. The key idea behind CH is to contract vertices in a total order, from low to high, by introducing shortcuts among their neighbors to maintain distance information. These shortcuts drastically reduce the search space during query time by allowing the algorithm to skip over intermediate nodes and directly access relevant paths, leading to faster query responses. Despite its efficiency in pruning the search space, CH may still require exploring many paths at query time, which can result in less than optimal performance. To address the limitations of search-based methods, labelling-based methods have been developed with great success [1, 2, 4, 5, 9, 18, 21, 32]. These methods precompute distance labels that capture the shortest-path distances between pairs of vertices. At query time, rather than performing a search over the graph, the algorithm simply examines the precomputed labels to retrieve the distances. Labelling-based methods can answer distance queries significantly faster than search-based methods, at the cost of requiring additional space for storing labels.

Despite the progress made in static road network algorithms, adapting these methods for dynamic road networks remains a significant challenge. A common approach is to incrementally update precomputed structures, such as shortcuts and distance labels, rather than recomputing them entirely from scratch. However, in dynamic settings, queries and updates naturally exhibit a trade-off when relying on pre-computed data structures to speed up performance. Search-based methods [14, 22, 27] focus on maintaining shortcuts, leading to faster updates but can result in significantly slower query times. Conversely, labeling-based methods provide fast query times by precomputing distance labels [9, 30, 32]. However, they face challenges with slow updates, as updating these labels in response to network changes is computationally demanding. The complexity of keeping the distance labels accurate makes labeling-based approaches less efficient for frequent updates. The inherent trade-off between query performance and update efficiency is a key challenge in dynamic road network algorithms.

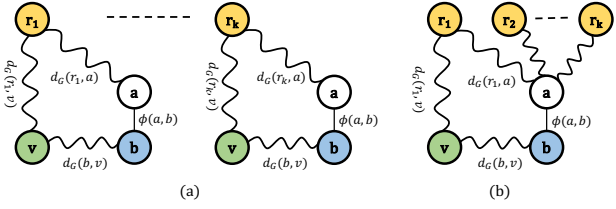

**Figure 1: An illustration of searches performed by our dynamic algorithms based on triangle inequality** ($d_G(r_i, a) + \phi(a,b)) + d_G(b,v) \leq d_G(r_i,v)$, **where** $i \in [1,k]$, $\phi(a,b)$ **is an update,** $v$ **is an affected vertex, and** $\{r_1, \ldots, r_k\}$ **is a set of ancestors: (a) Label searches, one from each ancestor; (b) Pareto searches combining multiple searches from ancestors.**

**Our Ideas.** In this work, we aim to develop an efficient solution for answering distance queries on dynamic road networks by addressing two key questions: (1) How to choose auxiliary data structures that balance query and update efficiency? (2) How to design algorithms that efficiently maintain these auxiliary data structures to reflect changes on dynamic road networks? We begin by analyzing *Hierarchical Cut 2-hop Labelling (HC2L)* [12], a recent method that achieves state-of-the-art results on static road networks. Despite its impressive performance, we observe that, due to the presence of shortcuts within partitions, its data structure introduces inefficiencies in maintaining distance queries in dynamic road networks. These shortcuts complicate the design of update mechanisms, making it difficult to efficiently modify the structure in response to real-time changes. To circumvent this inefficiency barrier, we do the following:

– We define *stable tree hierarchy* that exhibits several nice properties: (1) *structural stability*: a stable tree hierarchy is structurally independent of edge weights – an important condition for efficient maintenance [22, 27, 32]; (2) *balancedness*: a stable tree hierarchy is still balanced – inheriting from the balanced tree hierarchy of HC2L [12]; (3) *2-hop common ancestors*: Given any two vertices, every path between them contains at least one of their common ancestors. We propose the *Stable Tree Labelling* (STL), a labeling method built on a stable tree hierarchy. One novel and crucial design is that *labels only store distances within subgraphs, not across the entire graph*. This significantly reduces the number of labels affected by dynamic updates, thereby enhancing the efficiency of update operations.

– We propose algorithms to efficiently maintain STL from two different perspectives: one is *ancestor-centric*, namely *Label Search algorithms*, while the other is *update-centric*, namely *Pareto Search algorithms*. Label Search algorithms identify a set of *ancestors* that are sufficient to maintain STL and then perform an efficient search to update affected labels from each ancestor, as depicted in Figure 1(a). Nonetheless, searches from different ancestors may share common paths, e.g., paths between an affected vertex $v$ and the vertex $b$ incident to an update $(a,b,\phi)$ depicted in Figure 1(b). Based on this observation, *Pareto Search algorithms* improve Label Search algorithms by exploring the interaction between search spaces of different ancestors, and then combine searches from multiple ancestors into only two searches, eliminating duplicate search traversals.

Theoretically, we establish key properties to show the correctness of our Label Search and Pareto Search algorithms. We also derive complexity bounds for these two types of algorithms. Empirically, we evaluate our algorithms on 10 real-world large road networks, including the whole road network of USA and western Europe road network. The results show that our algorithms considerably outperform the state-of-the-art methods. For example, compared with IncH2H [32], our algorithms perform about *three times faster* in query time on all datasets, and *five to seven times faster* in update time on large road networks, while consuming an order of magnitude less space. Our algorithms are also several orders of magnitude faster than DTDHL [30] in terms of update time, while being significantly faster in terms of query processing and requiring only 25%-30% of space for labelling.

**Outline.** The rest of the paper is organized as follows. In Section 2, we discuss other works that are related to our work. In Section 3, we present basic notations and discuss state-of-the-art methods. In Section 4 we present our solution STL. We introduce two dynamic algorithms for edge weight decrease and increase, respectively, in Section 5 and analyze their time complexity in Section 6. Section 7 presents experimental results. Section 8 discusses the extensions to edge/node insertions/deletions and to directed road networks. Section 9 concludes the paper.

## 2 RELATED WORK

We review existing works for answering distance queries on dynamic road networks, which broadly fall into two categories: (1) *shortcut maintenance* – maintaining shortcuts used in search-based methods [14, 22, 27], and (2) *labelling maintenance* – maintaining distance labelling used in labelling-based methods [9, 30–32]. Below, we discuss each category in detail.

**Shortcut maintenance.** Geisberger et al.[14] proposed a vertex-centric algorithm that maintains *contraction hierarchy* (CH) [13] with minimal shortcuts. Their algorithm first finds vertices affected by dynamic changes and then recontracts these vertices to update affected shortcuts. This is highly inefficient because recontraction has to ensure the minimality of shortcuts - keeping only shortcuts that satisfy a shortest distance constraint [13]. Later on, some works [22, 27] followed similar ideas to maintain CH-W index [21] but without requiring the minimality requirement of shortcuts. By allowing redundant shortcuts, these methods can avoid insertion or deletion of shortcuts during maintenance and only update the weights of affected shortcuts. Accordingly, update time is improved at the cost of slower query time. However, since CH-W index may potentially create extremely dense structures on graphs with large treewidth, these approaches limits their applicability in practice.

**Labelling maintenance.** Following a different line of work, a dynamic algorithm, denoted as DynH2H, has been proposed [9], which maintains H2H [21] to answer distance queries efficiently under dynamic changes on road networks. Later, Zhang et al. [30] proposed an algorithm, called *dynamic tree decomposition based hub labelling* (DTDHL), which is an optimized version of DynH2H. As H2H-index constructs labels using CH-W index, DTDHL first updates shortcuts similar to DCH [22] and then updates labels via tree decomposition in the top-down manner. Recently, Zhang et al. [32] studied the theoretical boundedness of dynamic CH index [22, 27] and proposed IncH2H to maintain H2H index [9, 30] under edge weight increase and decrease. IncH2H has achieved the state-of-the-art performance for answering distance queries on dynamic road networks. However, it suffers from maintaining a huge index constructed based on CH-W index. As a result, IncH2H may contain a large number of distance entries for graphs with large treewidths. Together with auxiliary data used to speed up index updates, this can lead to huge memory requirements. For example, the index it maintains for the whole USA road network is over 300 GB in size.

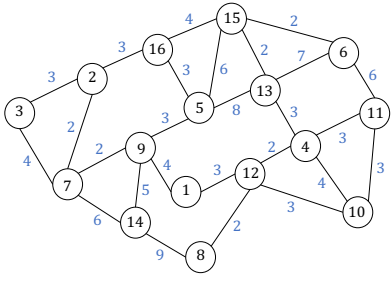

Figure 2: An example road network $G$.

## 3 PRELIMINARIES

Let $G = (V, E, \phi)$ be a road network where $V$ is a set of vertices, and $E$ is a set of edges. Each edge $(u, v) \in E$ is associated with a non-negative weight $\phi(u, v) \in \mathbb{R}_{\geq 0}$. A path is a sequence of vertices $p = (v_1, v_2, \ldots, v_k)$ where $(v_i, v_{i+1}) \in E$ for each $1 \leq i < k$. The weight of a path $p$ is defined as $\phi(p) = \sum_{i=1}^{k-1} \phi(v_i, v_{i+1})$. For two arbitrary vertices $s$ and $t$, a shortest path $p$ between $s$ and $t$ is a path starting at $s$ and ending at $t$ such that $\phi(p)$ is minimised. The distance between $s$ and $t$ in $G$, denoted as $d_G(s, t)$, is the weight of any shortest path between $s$ and $t$. We use $N_G(v)$ to denote the set of direct neighbors of a vertex $v \in V$, i.e. $N_G(v) = \{(u, \phi(u, v)) \mid u \in V, (u, v) \in E\}$, and $V(G)$ and $E(G)$ to refer to the set of vertices and edges in $G$, respectively. We consider two types of edge weight updates: increases and decreases. Table 1 summarizes the notations.

Table 1: Summary of Notations

| Notation | Description |
|---|---|
| $G = (V, E, \phi)$ | an undirected and weighted graph |
| $G \oplus \Delta G$ | applying a set of edge weight updates $\Delta G$ on $G$ |
| $N_G(v)$ | the set of neighbors of vertex $v$ in $G$ |
| $\mathcal{T}$ | a tree decomposition |
| $H$ | a balanced tree hierarchy |
| $T = (\mathcal{N}, \mathcal{E}, \ell)$ | a stable tree hierarchy |
| $\preceq$ | vertex partial order induced by $T$ |
| $\text{Anc}(v), \text{Desc}(v)$ | set of ancestors or descendants of vertex $v$ w.r.t. $\preceq$ |
| $G[\text{Desc}(v)]$ | the subgraph of $G$ induced by $\text{Desc}(v)$ |
| $d_G(v, u), d_G^w(v, u)$ | distance between $v$ and $u$ in $G$ or $G[\text{Desc}(w)]$ |
| $\tau(v)$ | the label index of a vertex $v$ (i.e., $|\text{Anc}(v)|$) |
| $L(v)$ | the label of a vertex $v$ |
| $L_v[r]$ | the distance from vertex $v$ to vertex $r$ in $L(v)$ |
| $\text{Lca}(v, u), \text{Ca}(v, u)$ | the (lowest) common ancestors of $v$ and $u$ in $T$ |
| $P_G(v, u), P_G^w(v, u)$ | shortest paths between $v$ and $u$ in $G$ or $G[\text{Desc}(w)]$ |

In the following, we first analyze *Incremental Hierarchical 2-Hop* (IncH2H) [32], the state-of-the-art method on dynamic road networks. Then, we present a recent method that achieves the state-of-the-art performance on static road networks, called *Hierarchical Cut 2-hop Labelling* (HC2L) [12], and discuss the limitations of extending it on dynamic road networks.

### 3.1 Incremental Hierarchical 2-Hop

*Incremental Hierarchical 2-Hop* (IncH2H) [32] maintains H2H-Index to answer distance queries on dynamic road networks.

<u>Construction.</u> H2H-Index is a 2-hop labelling constructed upon a vertex hierarchy, which is obtained via tree decomposition based on CH-W index [21]. Let $\pi$ be a total order on $V(G)$ and $G_S$ be the graph of CH-W index over $G$. H2H-Index first constructs a tree decomposition $\mathcal{T}$ by forming a tree node $X(v)$ for each contracted vertex $v \in V(G_S)$, which contains $v$ and all its neighbours $N_{G_S}(v)$ with shortcuts $\{(v, u) \mid u \in N_{G_S}(v)\}$. Then the tree node $X(u)$ of the vertex $u \in X(v) \setminus \{v\}$, where $u$ is the lowest ranked vertex in $X(v)$, is assigned as the parent of $X(v)$. This construction ensures that for each $v \in V(G)$, all vertices in $X(v)$ are its ancestors in $\mathcal{T}$. Another important property of $\mathcal{T}$ is that every shortest path between any two vertices $s$ and $t$ must pass through the lowest common ancestor of $s$ and $t$ in $\mathcal{T}$. Then, a 2-hop labelling is constructed using $\mathcal{T}$ such that the label $L(v)$ of each vertex $v \in V(G)$ consists of three arrays: (i) an *ancestor array* $[w_1, \ldots, w_k]$ representing the path from the root to $v$ in $\mathcal{T}$, (ii) a *distance array* $[\delta_{vw_1}, \ldots, \delta_{vw_k}]$ where $\delta_{vw_i} = d_G(v, w_i)$ and $\{w_1, \ldots, w_k\}$ is the set of vertices that are ancestors of $v$ in $\mathcal{T}$, and (iii) a *position array* $[i_1, \ldots, i_k]$ that stores positions of $[w_1, \ldots, w_k]$ in $\mathcal{T}$, where the position of $w_i$ is defined as its depth in $\mathcal{T}$.

<u>Maintenance.</u> H2H index is dynamically maintained in two phases: 1) *shortcut maintenance*, and 2) *labelling maintenance*. The shortcut maintenance phase identifies and updates the weights of affected shortcuts in $G_S$. The labelling maintenance phase updates affected labels in $L$ with the help of shortcut graph $G_S$. Affected shortcuts in $G_S$ are used to update the labels of affected vertices w.r.t. a set of ancestors. That is, for an affected shortcut $\langle u, v \rangle \in G_S$, it finds all ancestors $a$ whose distances to $u$ have been affected. Afterwards H2H-Index iteratively processes descendants of $u$ which are further affected by the changes in distances between $a$ and $u$.

<u>Querying.</u> Given any two vertices $s, t \in V(G)$ and their lowest common ancestor $\text{Lca}(s, t) = a$ in $\mathcal{T}$, $d_G(s, t)$ is computed as

$$d_G(s, t) = \min\{\delta_{sw_i} + \delta_{tw_i} \mid \delta_{sw_i} = L(s).dist(i), \quad (1)$$
$$\delta_{tw_i} = L(t).dist(i), i \in L(a).pos, a = \text{Lca}(s, t)\}.$$

*Example 3.1.* Figure 3 shows a tree decomposition $\mathcal{T}$ of $G$ depicted in Figure 2 and the labels of vertices $\{12, 11, 3\}$ in H2H-Index. $L(11)$ stores an ancestor array $[16, 15, 12, 5, 13, 11]$ containing all ancestors of 11, a distance array $[12, 8, 5, 14, 6, 0]$ storing the distances from vertex 11 to its ancestors, and a position array $[2, 3, 5, 6]$ which represents the positions of nodes $\{11, 13, 12, 15\}$ inside the tree node of vertex 11 in $\mathcal{T}$. Under edge weight updates, ancestor and position arrays remain intact while distance arrays are maintained. Suppose the weight of an edge $(1, 9)$ has increased, IncH2H first updates the weights of all affected shortcuts $\{(1, 9), (9, 12)\}$ in $G_S$ starting from the tree node of 1 in $\mathcal{T}$. Then, it iteratively identifies and updates affected labels of vertices $\{1, 2, 3, 5, 7, 8, 9, 10\}$ w.r.t. affected shortcuts $\{(1, 9), (9, 12)\}$. For a distance query between vertices 3 and 11, $\text{Lca}(3, 11) = 5$ is first obtained, and then using the distances in $L(3)$ and $L(11)$ at the positions $[1, 2, 3]$ in $L(5)$, $d_G(3, 11) = 18$ is obtained according to eq. (1).

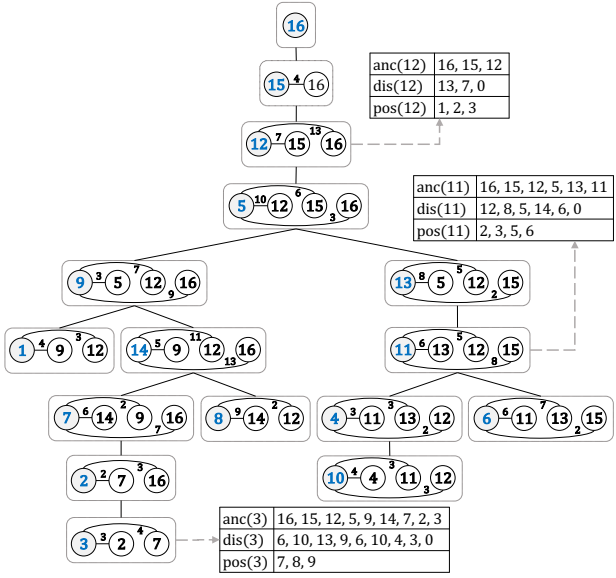

Figure 3: An illustration of H2H-Index.

## 3.2 Hierarchical Cut 2-Hop Labelling

Very recently, *Hierarchical Cut 2-hop Labelling* (HC2L) [12] was proposed, which exploits a vertex hierarchy by leveraging recursive balanced cuts on a road network and has shown to significantly outperform H2H-Index on static road networks.

*Construction.* Unlike H2H-Index, HC2L [12] develops a recursive algorithm to find balanced cuts that partition a road network into smaller components. The resultant cuts are arranged to form a balanced tree hierarchy which defines a vertex-quasi order $\preceq$ on $V(G)$. A balanced tree hierarchy $H$ over $G$ has the nice property that each internal node of $H$ is a separator between its left and right subtrees. This allows to leverage the least common ancestor of two vertices $s$ and $t$ in $H$ to find vertices that separates them. A 2-hop labelling $L$ is constructed upon $H$ by computing $L(v) = [\delta_{vw_1}, \ldots, \delta_{vw_k}]$ for each $v \in V(G)$, where each $\delta_{vw_i} = d_G(v, w_i)$ denote the distance to $v$ from its ancestors $\{w_1, \ldots, w_k\}$.

*Querying.* Given any two vertices $s, t \in V(G)$ and $\text{Lca}(s,t)$ in $H$, the distance between $s$ and $t$ is computed as the minimum value of distances stored in $L(s)$ and $L(t)$ to vertices in $\text{Lca}(s,t)$ as

$$d_G(s,t) = min\{\delta_{sr} + \delta_{tr} \mid \qquad (2)$$
$$\delta_{sr} \in L(s), \delta_{tr} \in L(t), r \in \text{Lca}(s,t)\}.$$

*Example 3.2.* Figure 4 illustrates a balanced tree hierarchy along with the labels of vertices $\{9, 14, 11, 16\}$ in HC2L for $G$ shown in Figure 2. The distance between two vertices 11 and 16 can be obtained via the $\text{Lca}(11, 16) = \{15\}$. The level 1 of $\text{Lca}(11, 16)$ is first computed using bitstrings 1000 of 11 and 11 for 16. Then using the cut distances stored at level 1 in $L(11)$ and $L(16)$, $d_G(11, 16) = 12$.

*Discussion.* Despite achieving the state-of-the-art performance on dynamic road networks, IncH2H has drawbacks. It constructs a tree decomposition based on CH-W index [22], which often leads to a large height and width. Consequently, the index size of IncH2H can be huge which may hinder IncH2H to efficiently perform maintenance. IncH2H also requires a complex mechanism for computing the least common ancestor of two vertices, which degrades query performance. In contrast, HC2L exploits balanced tree structures, significantly outperforming IncH2H for answering distance queries on static road networks. However, on dynamic road networks H2CL has a major drawback. Since HC2L adds shortcut edges to ensure the preservation of distances when constructing a balanced tree hierarchy, maintaining such a balanced tree hierarchy incrementally requires shortcut edges to be added (or removed). This would make cuts at the lower levels of a balanced tree hierarchy no longer vertex separators, and large portions of the balanced tree hierarchy and the labels have to be reconstructed. As a result, maintaining a balanced tree hierarchy to reflect dynamic changes on $G$ is expensive.

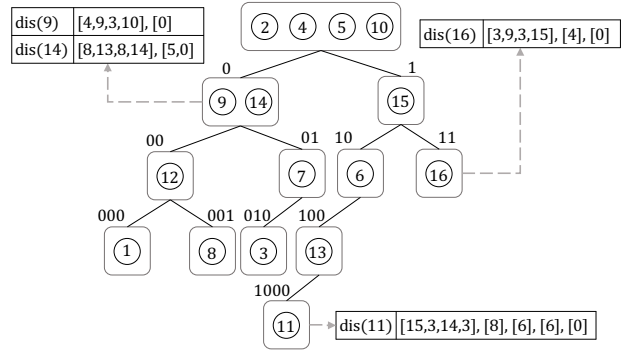

Figure 4: An illustration of HC2L framework.

## 4 STABLE TREE LABELLING

In this section, we present a simple yet efficient labelling method which alleviates limitations of IncH2H to support fast query processing and maintenance on dynamic road networks.

**Stable Tree Hierarchy.** Below, we define a tree hierarchy over $V(G)$ without any shortcuts. This eliminates expensive maintenance of shortcuts for dynamic changes and significantly reduces construction time. Compared with [12], the omission of shortcuts in our work leads to smaller cuts at lower levels as the subgraphs remain sparse, accordingly reducing both the number of common ancestors and overall labelling size.

*Definition 4.1 (Stable Tree Hierarchy).* A *stable tree hierarchy* is a binary tree $T = (\mathcal{N}, \mathcal{E}, \ell)$, where $\mathcal{N}$ is a set of tree nodes, $\mathcal{E}$ is a set of tree edges, and $\ell : V(G) \to \mathcal{N}$ is a total surjective function, satisfying the following conditions:

(1) Each $N \in \mathcal{N}$ satisfies

$$|T_\downarrow(N_l)|, |T_\downarrow(N_r)| \leq (1-\beta) \cdot |T_\downarrow(N)|$$

where $0 < \beta \leq 0.5$, $T_\downarrow(N)$ denotes a subtree rooted at $N$, and $N_l$ and $N_r$ are the left and right children of $N$, respectively.

(2) For any two vertices $s, t \in V(G)$, the following is satisfied:

$$p \in P_G(s,t) \implies V(p) \cap \text{Ca}(s,t) \neq \emptyset$$

where $P_G(s,t)$ is the set of all shortest paths between $s$ and $t$ in $G$, and $\text{Ca}(s,t)$ is the set of vertices in all common ancestors of $\ell(s)$ and $\ell(t)$ in $T$.

*Example 4.2.* Consider two vertices 11 and 13 in $G$ shown in Figure 2 and the corresponding stable tree hierarchy $T$ shown in Figure 5(a). We have $\text{Ca}(11, 13) = \{2, 4, 5, 10, 6\}$ and $\langle 11, 4, 13 \rangle \in P_G(s,t)$ contains a vertex $4 \in \text{Ca}(11, 13)$.

**Labelling Construction.** A stable tree hierarchy defines a partial order between tree nodes, which can be expanded to vertices by imposing an arbitrary total order between vertices associated with the same tree node.

*Definition 4.3 (Vertex Partial-Order).* Let $T$ be a stable tree hierarchy, and $\preceq_t$ an arbitrary total order on $V$. Then $w \preceq v$ iff $\ell(w)$ is a strict ancestor of $\ell(v)$, or $\ell(w) = \ell(v)$ and $w \preceq_t v$.

Given any vertex $v \in V(G)$, the *ancestors* of $v$ w.r.t. $\preceq$ is the set of all preceding vertices, i.e., $\text{Anc}(v) = \{w \in V(G) \mid w \preceq v\}$.

*Definition 4.4 (Label Index).* Let $v \in V(G)$. The *label index* $\tau(v)$ of vertex $v$ is the position of $v$ w.r.t. $\preceq$, i.e., $\tau(v) = |\text{Anc}(v)|$.

*Example 4.5.* Consider Figure 5(a) again, the label index $\tau(5)$ of vertex 5 is 2 because there are 2 vertices $\{2, 4\}$ preceding vertex 5. Similarly, the label index of vertex 12 is 6 because there are 6 vertices $\{2, 4, 5, 10, 9, 14\}$ preceding vertex 12.

Let $d_G^w(v, u)$ denote the distance between $v$ and $u$ in the subgraph of $G$ induced by $\text{Desc}(w) = \{x \in V(G) \mid w \preceq x\}$.

*Definition 4.6 (Stable Tree Labelling).* Let $T$ be a stable tree hierarchy over $G$. A *stable tree labelling* (STL) over $T$ is a distance labelling $L = \{L(v) \mid v \in V(G)\}$ where the label $L(v)$ of each vertex $v$ is defined as a distance array $L(v) = [\delta_{vw_1}, \ldots, \delta_{vw_k}]$, with $\text{Anc}(v) = \{w_1, \ldots, w_k\}$, $w_1 \preceq \ldots \preceq w_k$, and $\delta_{vw_i} = d_G^{w_i}(v, w_i)$.

Unlike prior work, distances stored in our labels are not distances in $G$, but distances within subgraphs. This restriction simplifies not only label construction but also label updates. In particular, a label can only be affected by an edge weight update if that edge lies in the relevant subgraph, and thus fewer labels need to be updated. Despite this, stable tree labellings satisfy the 2-hop cover property.

LEMMA 4.7. *For any vertices $s, t \in V(G)$, there exists at least one vertex $r \in \text{Anc}(s) \cap \text{Anc}(t)$ and distance entries $\delta_{sr} \in L(s)$ and $\delta_{tr} \in L(t)$ such that $\delta_{sr} + \delta_{tr} = d_G(s, t)$.*

PROOF. Let $p \in P_G(s, t)$ and $r$ be the vertex in $p$ with the minimal label index $\tau(r)$. Then $r \preceq v$ for all $v \in V(p)$ by Definition 4.4. Thus $p$ lies in $\text{Desc}(r)$. It follows that $\delta_{sr} + \delta_{tr} = \phi(p) = d_G(s, t)$. □

**Distance Queries.** A distance query $Q(s, t)$ is answered as:
$$d_G(s, t) = \min\{\delta_{sr} + \delta_{tr} \mid \delta_{sr} \in L(s), \delta_{tr} \in L(t),$$
$$r \in \text{Anc}(s) \cap \text{Anc}(t)\}. \quad (3)$$

Using all common ancestors as hubs can make query answering more expensive, especially for local queries. However, as label entries of common ancestors are stored consecutively in memory, this leads to highly efficient caching and avoids extra work associated with looking up which label entries to compare. We can quickly find $\text{CA}(s, t)$ using the level of $\text{LCA}(s, t)$ in the stable tree hierarchy. As in [12], we compute the level of $\text{LCA}(s, t)$ via bitstrings in $O(1)$ time, specifically as the length of the common prefix of the bitstrings of $s$ and $t$. The distance pairs used in eq. (3) can then be found at levels less than or equal to the level of $\text{LCA}(s, t)$.

*Example 4.8.* Consider a distance query $Q(11, 16)$ on $G$ shown in Figure 2. The bitstrings of vertices 11 and 16 are 11 and 100, respectively, shown in Figure 5(a). The level $l$ of $\text{LCA}(11, 16)$ is 1 and $\text{CA}(11, 16) = \{2, 4, 5, 10, 6\}$ at levels $0 \leq l \leq 1$, $d_G(11, 16) = 12$ is obtained using $L(11)$ and $L(16)$ w.r.t. $\text{CA}(11, 16)$ in Equation 3.

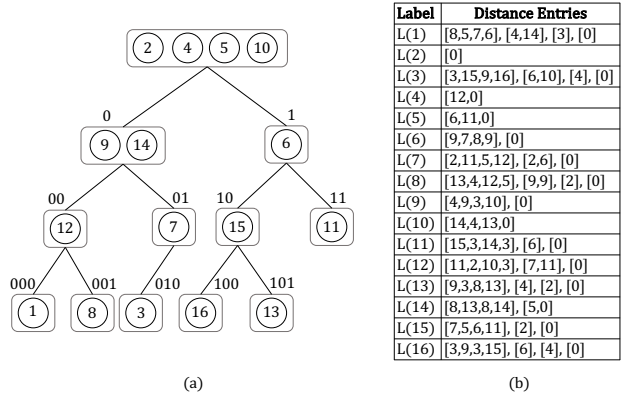

Figure 5: Stable Tree Hierarchy $T$ and Labelling $L$.

*Remark* 1. To construct stable tree hierarchies, we modify the recursive bi-partitioning algorithm presented in [12] to avoid the addition of shortcuts for distance preservation. We compute a partition bitstring for each cut $N$ and compute label index $\tau(r)$ and distance $d_G^r(v, r)$ within the subgraph $G[\text{Desc}(r)]$ for each $r \in N$ and $v \in \text{Desc}(r)$. Thus, by omitting shortcuts, our labels store only distances within subgraphs, which suffices by Lemma 4.7. Furthermore, omitting shortcuts results in smaller cuts, reducing both number of common ancestor vertices and overall labeling size.

## 5 DYNAMIC ALGORITHMS

In this section, we present two efficient algorithms to maintain stable tree labelling: one is called *Label Search algorithm* and the other is called *Pareto Search algorithm*. The key idea of these algorithms is to identify affected labels to update by performing pruned searches w.r.t. ancestors starting from an updated edge using triangle inequality illustrated in Figure 1.

*Definition 5.1 (Affected Vertex).* A vertex $v$ is *affected* w.r.t. an ancestor $r$ by an update on edge weight iff the set of shortest paths between $v$ and $r$ or their length changes.

LEMMA 5.2. *Let $(a, b, \phi_{old})$ and $(a, b, \phi_{new})$ be an edge before and after the update. If $\phi_{old} < \phi_{new}$ then $v$ is affected w.r.t. $r$ iff $d_G(r, v) = d_G(r, a) + \phi_{old} + d_G(b, v)$. If $\phi_{old} > \phi_{new}$ then $v$ is affected w.r.t. $r$ iff $d_G(r, v) \geq d_G(r, a) + \phi_{new} + d_G(b, v)$.*

Let $\Delta G$ be a set of edge weight updates on $G$. For clarity, we also use $L_v[r]$ to refer to the distance from $v$ to an ancestor $r \in \text{Anc}(v)$ stored in the label $L(v)$, i.e., $L_v[r] = \delta_{vr}$ where $\delta_{vr} \in L(v)$.

### 5.1 Label Search Algorithm

The idea of Label Search algorithm is to perform a single search for updates w.r.t. each ancestor of affected vertices. Algorithms 1 and 2 describe the steps for weight decrease and increase, respectively.

The lemma below states that, for two vertices incident to any edge, one must precede the other in a stable tree hierarchy.

LEMMA 5.3. *Let $T$ be a stable tree hierarchy on $G$. If $(u, v) \in E(G)$, then either $u \preceq v$ or $v \preceq u$ hold on $T$.*

*5.1.1 Edge Weight Decrease.* For the decrease case, by Lemma 5.3, Algorithm 1 first partitions updates $(a, b, \phi_{new}) \in \Delta G$ w.r.t. each ancestor $r \in \text{Anc}(a)$ and push them to their corresponding priority queue $Q_r$ (Lines 2-7). Then a search w.r.t. $Q_r$ progresses

**Algorithm 1:** Label Search (Decrease)

1 **Function** Search-and-Repair($L, T, \Delta G$)
2    **foreach** $(a, b, \phi_{new}) \in \Delta G$ with $\tau(b) > \tau(a)$ **do**
3       **foreach** $r \in [0, \tau(a)]$ **do**
4          **if** $L_a[r] + \phi_{new} < L_b[r]$ **then**
5             add $(L_a[r] + \phi_{new}, b)$ into $Q_r$
6          **else if** $L_b[r] + \phi_{new} < L_a[r]$ **then**
7             add $(L_b[r] + \phi_{new}, a)$ into $Q_r$

8    **foreach** $Q_r$ **do**
9       **foreach** $(d, v) \in Q_r$ in increasing order of $d$ **do**
10          **if** $d < L_v[r]$ **then**
11             $L_v[r] \leftarrow d$
            // visit neighbors
12             **foreach** $(n, \phi_n) \in N(v)$ with $\tau(n) > \tau(r)$ **do**
13                **if** $d + \phi_n < L_n[r]$ **then**
14                  add $(d + \phi_n, n)$ to $Q_r$

---

**Algorithm 2:** Label Search (Increase)

1 **Function** Search($L, T, \Delta G$)
2    **foreach** $(a, b, \phi_{old}) \in \Delta G$ with $\tau(b) > \tau(a)$ **do**
3       **foreach** $r \in [0, \tau(a)]$ **do**
4          **if** $L_a[r] + \phi_{old} = L_b[r]$ **then**
5             add $(L_a[r] + \phi_{old}, b)$ into $Q_r$
6          **else if** $L_b[r] + \phi_{old} = L_a[r]$ **then**
7             add $(L_b[r] + \phi_{old}, a)$ into $Q_r$

8    **foreach** $Q_r$ **do**
9       **foreach** $(d, v) \in Q_r$ in increasing order of $d$ **do**
10          **if** $v \notin V_{\text{AFF}}$ **then**
11             add $v$ into $V_{\text{AFF}}$
            // visit neighbors
12             **foreach** $(n, \phi_n) \in N(v)$ with $\tau(n) > \tau(r)$ **do**
13                **if** $d + \phi_n = L_n[r]$ **then**
14                  add $(d + \phi_n, n)$ to $Q_r$

15       Repair($r, V_{\text{AFF}}$)

16 **Function** Repair($r, V_{\text{AFF}}$)
17    **foreach** $v \in V_{\text{AFF}}$ **do**
18       $L_v[r] \leftarrow \infty$
19       $L_v[r] \leftarrow \min \left\{ L_n[r] + \phi \;\middle|\; \begin{array}{l}(n, \phi) \in N(v) \setminus V_{\text{AFF}} \\ \text{with } \tau(n) > \tau(r)\end{array}\right\}$
20       **if** $L_v[r] \neq \infty$ **then**
21          add $(L_v[r], v)$ into $Q_r$

22    **foreach** $(d, v) \in Q_r$ in increasing order of $d$ **do**
23       **if** $d < L_v[r]$ **then**
24          $L_v[r] \leftarrow d$
25          **foreach** $(n, \phi_n) \in N(v)$ with $\tau(n) > \tau(r)$ **do**
26             **if** $d + \phi_n < L_n[r]$ **then**
27                add $(d + \phi_n, n)$ to $Q_r$

---

as follows. Starting from affected vertices in $Q_r$ that are incident to updated edges, we repeatedly explore their neighbors that have a larger label index than $r$ in order to find affected vertices whose distances to $r$ are changed using Lemma 5.2 (Lines 10-12). As edge weight decrease only shrinks lengths of the shortest-paths, the new distance $d$ from an affected node $v$ to $r$ becomes known when it is processed at Line 7, which allows us to immediately update its label if the length of a newly found path is shorter than the existing one at Line 9.

Two important aspects contribute to the efficiency of the above process: (1) $Q_r$ processes vertices in increasing order of their distances to $r$, ensuring that each vertex is processed at most once. (2) Only the label index $\tau(r)$ is required when looking up $L_v[r]$.

*5.1.2 Edge Weight Increase.* For the increase case, Algorithm 2 also groups updates $\Delta G$ into the priority queues $Q_r$ w.r.t. each ancestor $r \in \text{Anc}(a)$ (Lines 2-7). Then starting from $Q_r$, we iteratively visit neighbors with label index larger than $r$ to identify a set of affected vertices $V_{\text{AFF}}$ (Lines 6-12). However, different from the decrease case, we cannot repair the label of an affected vertex immediately, as its new distance value to $r$ is unknown at Line 7. Instead, we employ an efficient repairing mechanism (i.e., Function Repair) which repairs the labels for all affected vertices once in only one go. Specifically, we first compute distance bounds of vertices in $V_{\text{AFF}}$ w.r.t. $r$ using their unaffected neighbors that have a label index larger than $r$ (Line 17).

**Definition 5.4 (Distance Bound).** Let $V_{\text{AFF}} \subseteq \{u \in V(G) \mid u \prec r\}$ and $v \in V_{\text{AFF}}$. The *distance bound* of $(v, r)$ w.r.t. $V_{\text{AFF}}$ is:

$$d(v, r, V_{\text{AFF}}) := \min \left\{ L_u[r] + \phi \;\middle|\; \begin{array}{l}(u, \phi) \in N(v) \setminus V_{\text{AFF}} \\ \text{and } \tau(u) > \tau(r)\end{array}\right\}$$

The following lemma allows us to compute the distance of vertices in $V_{\text{AFF}}$ from ancestor $r$ using their distance bounds.

**Lemma 5.5.** *Let $V_{\text{AFF}} \subseteq \{u \in V \mid u \prec r\}$ and $v \in V_{\text{AFF}}$ with minimal distance bound. Then $L_v[r] = d(v, r, V_{\text{AFF}})$.*

We enqueue the affected vertices with finite distance bounds into $Q_r$ (Lines 18-19). Then we start processing vertices in $Q_r$ in increasing order of their distance bounds and repeatedly enqueue their affected neighbors (Lines 23-25) into the queue.

*Example 5.6.* Suppose that the weight of an edge $(1, 9)$ in Figure 2 is decreased from 4 to 1. The set of ancestors to which distances need an update are $\{2, 4, 5, 10, 9\}$ as $\tau(9) < \tau(1)$. Accordingly, the priority queues w.r.t. these ancestors are $Q_2 = \{(5, 1)\}, Q_4 = \{(6, 9)\}, Q_5 = \{(4, 1)\}, Q_{10} = \{(7, 9)\}$ and $Q_9 = \{(1, 1)\}$. Figure 6(a) illustrate searches from $Q_2$ and $Q_5$, which start at the first affected vertex 1 highlighted in blue, and iteratively find and repair the labels of 6 affected vertices $\{1, 4, 8, 10, 11, 12\}$. The old and new distances w.r.t. ancestors $\{2, 5\}$ are shown next to each affected vertex in the form "*old → new*". Note that the label of vertex 4 does not store distance to 5, as 5 is not an ancestor of 4, and the distance to 5 at vertex 11 is updated to 13 as our search is restricted to $G[\text{Desc}(5)]$.

Now consider the weight of $(1, 9)$ is increased from 4 to 7, the set of ancestors remains the same, with the priority queues changed to $Q_2 = \{(8, 1)\}, Q_4 = \{(9, 9)\}, Q_5 = \{(7, 1)\}, Q_{10} = \{(10, 9)\}$ and $Q_9 = \{(4, 1)\}$. The searches from $Q_2$ and $Q_5$, illustrated in Figure 6(b)-(c), first mark vertices $\{1, 8, 10, 12\}$ as affected in Figure 6(b), and then repair their distances w.r.t. $\{2, 5\}$ in Figure 6(c). In Figure 6(c), the distance from 2 to vertex 10 is set to 16, as the previous shortest-path passing through the updated edge $(1, 9)$ is no longer the shortest one. Note again that we do not update the distance between 5 and 10 to 15, the distance in $G$, as our index stores distances in subgraphs, and ancestor 4 lies outside of $G[\text{Desc}(5)]$.

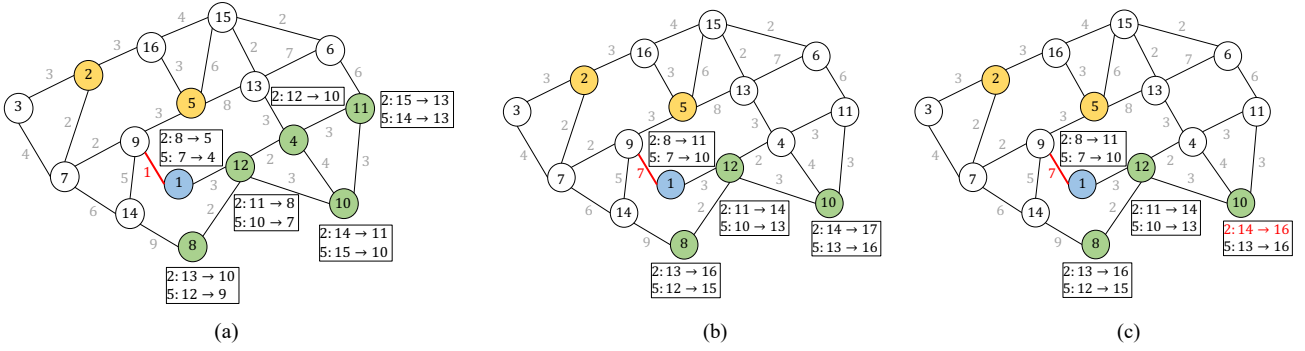

Figure 6: Searches w.r.t. ancestors {2, 5} performed by our algorithms: (a) searches for an edge weight decrease, and (b) & (c) searches for an edge weight increase, where yellow vertices denote ancestors, green vertices denote those being affected/repaired, and blue vertices denote the starting point of the searches.

## 5.2 Pareto Search Algorithm

*Observation.* The Label Search algorithm performs many searches from different ancestors. This may lead to common sub-paths being traversed multiple times. The Pareto Search algorithm aims to eliminate duplicate searches by combining searches from all ancestors into only two searches, starting from the two endpoints of an updated edge. Specifically, for an update $(a, b, \phi_{new})$, the distance between an ancestor $r$ and vertex $v$ changes if we have $d(r, v) = d(r, a) + d(a, v)$ or $d(r, v) = d(r, b) + d(b, v)$, depending on whether $r$ is closer to $a$ or $b$. Suppose that $a$ is closer to $r$. Since the distance from $r$ to $a$ is already stored in the label $L_a[r]$ and does not change, we only need to compute the new distance from $a$ to $v$ once for all ancestors and repair $L_v[r]$ accordingly.

*Example 5.7.* Consider a weight decrease of edge (1, 9) in Figure 2 by one. As the edge (1, 9) lies on the shortest paths from ancestors 2 and 5 to nodes {1, 8, 10, 12}, the update will affect the distances stored between them. It also lies on the shortest paths from ancestors 4 and 10 to nodes {3, 7, 9, 14}. Consider now the shortest path trees rooted at 2 and 5 to affected nodes {1, 8, 10, 12}, shown in Figure 7 – their subtrees rooted at 9 are identical. The same holds for the shortest path trees rooted at 4 and 10 to nodes {3, 7, 9, 14} and their subtrees rooted at 1. Next, we compute the new distances between 9 and nodes {1, 8, 10, 12} starting from vertex 9. Distances from ancestors 2 and 5 to those nodes are then computed by adding their distances from 9 to the distances from 2 or 5 to 9, respectively. The latter are stored in the label of vertex 9.

Unfortunately, things are not quite as simple as the last two examples suggest. Recall that our labels do not store distances in the entire graph $G$, but distances in subgraphs of $G$ – specifically, $\delta_{vw} \in L(v)$ is the distance between $v$ and $w$ in $G[\text{Desc}(w)]$. This restriction simplifies distance computation during index construction. However, it also means that ancestors of lower partial vertex order may follow paths that ancestors of higher partial vertex order may not, which makes it difficult to combine searches.

*Example 5.8.* Consider again Example 5.7, except that this time the weight of edge (1,9) is decreased all the way down to 1, as shown in Figure 6(a). Thus, the distances from ancestors {2, 5} to vertices {4, 11} change as well. As $2 \preceq 4 \preceq 5$, the search from 2 must consider paths passing through 4, while the search from 5 must not. The shortest paths (starting from root 9) w.r.t. to ancestors 2 and 5 are ⟨9, 1, 12, 4, 11⟩ and ⟨9, 1, 12, 10, 11⟩, respectively. In this case, we cannot simply combine the searches from 2 and 5 into a single search rooted at vertex 9.

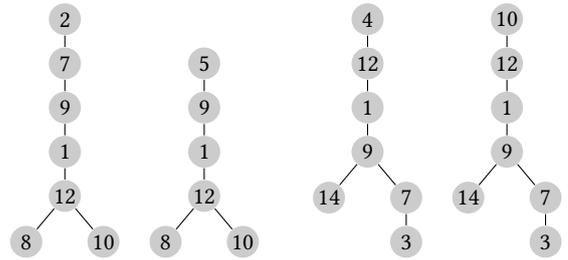

Figure 7: Search trees from Example 5.7

*Pareto Search.* To address this, we first observe that the subgraphs associated with different ancestors form an inclusion chain, meaning we have $S_a \subseteq S_b$ iff $b \preceq a$, where $S_a$ refers to the subgraph $G[\text{Desc}(a)]$. We will also use the notation $S_{\tau(a)}$ where this does not cause confusion. Hence the sequence of $d_{S_i}(u, v)$ values for distances between $u$ and $v$ w.r.t. $S_i$ is monotonically increasing. However, this increase is not *strictly* monotonic, i.e., there can be intervals of $i$ values with the same distance.

LEMMA 5.9. *For vertices $u, v, s, t \in V(G)$ with $u \preceq v \preceq s$ and $u \preceq v \preceq t$, we have $d_{S_u}(s, t) \leq d_{S_v}(s, t)$.*

*Example 5.10.* Consider again Example 5.8. We have $2 \preceq 4 \preceq 5 \preceq 10$ with $\tau(2) = 0, \tau(4) = 1, \tau(5) = 2$ and $\tau(10) = 3$. As $S_3 \subset S_2 \subset S_1 \subset S_0$ we must have $d_{S_0}(9, v) \leq d_{S_1}(9, v) \leq d_{S_2}(9, v) \leq d_{S_3}(9, v)$ for all $v \in S_3$, and in principle those values could all be different. However, $d_{S_0}(9, 11) = d_{S_1}(9, 11) = 9$ as the shortest path from 9 to 11 in $S_0 = G$ does not pass through vertex 2 excluded from $S_1$, and $d_{S_2}(9, 11) = d_{S_3}(9, 11) = 10$ as the shortest path from 9 to 11 in $S_2 = G \setminus \{2, 4\}$ does not pass through 5.

Let $u$ and $v$ be two vertices with $u \preceq v$. We use $(d, i)$ to denote the distance of $u$ and $v$ in the subgraph $S_i$, i.e., $d = d_{S_i}(u, v)$ and $i = \tau(u)$. Running separate searches for each subgraph $S_i$ means tracking $(d, i)$ values for all $i \in [0, \tau(v)]$, unless pruned. We improve on this by tracking only the Pareto-optimal pairs $(d, i)$ as defined below.

## Algorithm 3: Pareto Search (Decrease)

1 **Function** ParetoDec($a, b, \phi_{new}$)
2     Search-and-Repair ($a, b, \phi_{new}$)
3     Search-and-Repair ($b, a, \phi_{new}$)
4 **Function** Search-and-Repair($r, r', \phi$)
5     **foreach** $v \in V$ **do**
6        $level(v) \leftarrow 0$ // next level to process
    // init queue to start search from $r$ at $r'$
7     $r_{\min} \leftarrow \min(\tau(r), \tau(r'))$
8     add $(\phi, r', [0, r_{\min}])$ to $Q$
    // Dijkstra with intervals
9     **foreach** $(d, v, I_{active}) \in Q$ in order **do**
10       $I_{active}.max \leftarrow \min(I_{active}.max, \tau(v))$
11       $I_{active}.min \leftarrow \max(I_{active}.min, level(v))$
12       **if** $I_{active}.min > I_{active}.max$ **then**
13          **continue**
14       $level(v) \leftarrow I_{active}.max + 1$
      // update distance labels & find new active interval
15       **foreach** $i \in I_{active}$ **do**
16          **if** $d + L_r[i] < L_v[i]$ **then**
17             $L_v[i] \leftarrow d + L_r[i]$
18             **if** $min = $ undefined **then**
19                 $min \leftarrow i$
20             $max \leftarrow i$
      // visit neighbors
21       **if** $min \neq $ undefined **then**
22          **foreach** $(n, \phi_n) \in N(v)$ **do**
23             add $(d + \phi_n, n, [min, max])$ to $Q$

## Algorithm 4: Pareto Search (Increase)

1 **Function** ParetoInc($a, b, \phi_{old}, \phi_{new}$)
2     $\Delta = \phi_{new} - \phi_{old}$
3     Search ($a, b, \phi_{old}$)
4     Search ($b, a, \phi_{old}$)
5 **Function** Search($r, r', \phi$)
6     **foreach** $v \in V$ **do**
7        $level(v) \leftarrow 0$ // next level to process
    // init queue to start search from $r$ at $r'$
8     $r_{\min} \leftarrow \min(\tau(r), \tau(r'))$
9     add $(\phi, r', [0, r_{\min}])$ to $Q$
    // Dijkstra with intervals
10     **foreach** $(d, v, I_{active}) \in Q$ in order **do**
11       $I_{active}.max \leftarrow \min(I_{active}.max, \tau(v))$
12       $I_{active}.min \leftarrow \max(I_{active}.min, level(v))$
13       **if** $I_{active}.min > I_{active}.max$ **then**
14          **continue**
15       $level(v) \leftarrow I_{active}.max + 1$
      // update distance labels & find new active interval
16       **foreach** $i \in I_{active}$ **do**
17          **if** $d + L_r[i] = L_v[i]$ **then**
18             $L_v[i] \leftarrow L_v[i] + \Delta$
19             **if** $min = $ undefined **then**
20                 $min \leftarrow i$
21             $max \leftarrow i$
      // visit neighbors
22       **if** $min \neq $ undefined **then**
23          **foreach** $(n, \phi_n) \in N(v)$ **do**
24             add $(d + \phi_n, n, [min, max])$ to $Q$
         // update range of affected ancestors
25          **if** $affected(v).min = $ undefined **then**
26             $affected(v).min = min$
27          $affected(v).max = max$
28     Repair(*affected*)

*Definition 5.11 (Pareto-Optimal Pairs).* A pair $(d, i)$ is said to be *Pareto-optimal* iff there does not exist any other pair $(d', i')$ satisfying $d' \leq d$ and $i \geq i'$ and $(d, i) \neq (d', i')$.

*Example 5.12.* Consider again Example 5.10, where edge (9,1) in Figure 2 is updated to weight one. The $(d, i)$ pairs with minimal distance values $d$ for $S_0, \ldots, S_3$ at $v = 11$ are (9, 0), (9, 1), (10, 2) and (10, 3). Amongst these, (9, 1) and (10, 3) are Pareto-optimal.

Computing Pareto-optimal pairs essentially means to combine searches for ancestors with different $i$ but the same $d$. Ancestors are thus combined form intervals, which depend on the visited node $v$. E.g. for $v = 12$ in Example 5.8, distances in $S_0$ and $S_2$ are identical, but for $v = 11$ they differ.

*Proposed Algorithm.* During Pareto searches, we use a priority queue of $(d, i, v)$ tuples instead of using a priority queue of $(d, v)$ pairs. Here $i$ denotes the minimal value $\tau(w)$ of any node $w$ that the path being tracked passed through, so that $d$ describes the length of a path in $S_i$. Processing $(d, i, v)$ tuples with minimal $d$ value first ensures that paths found are minimal, as for standard Dijkstra, while breaking ties to process $(d, i, v)$ tuples with maximum $i$ value first ensures that Pareto-optimal tuples are encountered before others with the same $d$ (and $v$) value. By storing the smallest $i$ value that is not yet processed for each vertex, we can easily identify and discard tuples that are not Pareto-optimal.

Let $level(v)$ be the maximum $i$ value processed for a vertex $v$ in order to prune tuples that are not Pareto-optimal. Tracking $level(v)$ allows us to identify the interval of subgraphs for which the distance value $d$ of $(d, i, v)$ being processed is minimal as $[level(v), i]$, referred to as *Pareto-active interval* of $v$.

*Example 5.13.* Consider Example 5.10. We start our Pareto Search from vertex 9 by adding $(0, \tau(9) = 4, 9)$ into the queue. When visiting vertex 4, we update $i$ to $\tau(4) = 1$ and enqueue $(9, 1, 11)$. Since the tuple passing through vertex 10 has $i$ updated to $\tau(10) = 3$, $(10, 3, 11)$ is enqueued. Of those two, $(9, 1, 11)$ is processed first. After $(9, 1, 11)$ is processed, we set $level(11)$ to $1 + 1 = 2$ as the highest $i$ value processed at vertex 11 so far is 1. Any tuple $(d', i', 11)$ with $i' < level(11)$ popped from the queue afterwards can simply be discarded, knowing that any tuple $(d, i, 11)$ with $i \geq i'$ and $d \leq d'$ has been processed earlier. Thus, the discarded tuple is either not Pareto-optimal or a duplicate. When the second tuple $(10, 3, 11)$ is processed, $level(11)$ is set to $3 + 1 = 4$. The Pareto-active interval of vertex 11 was $[0, 1]$ first and then $[2, 3]$.

To prune vertices whose distances remain unaffected from our search, we check each level in the Pareto-active interval $[min, max]$ of a vertex $v$ being visited, following Lemma 5.2. Specifically, for each level $i \in [min, max]$, we compare distance

**Algorithm 5:** Pareto Repair (Increase)

1 **Function** Repair(*affected*)
  // initialize queue
2   **foreach** $v$ with *affected*($v$) $\neq$ undefined **do**
3     **foreach** $(n, \phi_n) \in N(v)$ **do**
4       **foreach** $i \in$ *affected*($v$) with $i \leq \tau(n)$ **do**
5         **if** $L_n[i] + \phi_n < L_v[i]$ **then**
6           add $(L_n[i] + \phi_n, v, i)$ to $Q$

  // repair
7   **foreach** $(d, v, i) \in Q$ in increasing order of $d$ **do**
8     **if** $d < L_v[i]$ **then**
9       $L_v[i] \leftarrow d$
10       **foreach** $(n, \phi_n) \in N(v)$ with $i \in$ *affected*($n$) **do**
11         **if** $L_v[i] + \phi_n < L_n[i]$ **then**
12           add $(L_v[i] + \phi_n, n, i)$ to $Q$

value $d + L_b[i]$ with $L_v[i]$ where $d$ is the length of the search path from vertex $b$ (root) to vertex $v$, while $L_b[i] = d_{S_i}(r, b)$ and $L_v[i] = d_{S_i}(r, v)$ represent distances from $b$ and $v$ to the ancestor $r$ at level $i$, respectively, stored in the labels. For weight decrease, we prune if $d + L_b[i] \geq L_v[i]$, while for weight increase, we prune if $d + L_b[i] < L_v[i]$. As we may only be able to prune a vertex w.r.t. some levels in its Pareto-active interval, we store the minimal interval containing all active levels of $v$, which we refer to as the *active interval* $I_{\text{active}}$ in Algorithms 3-4.

Our Pareto Search algorithm for edge weight decrease is presented in Algorithm 3. We perform two Pareto Searches starting from the two endpoints of the updated edge $(a, b, \phi_{new})$ as described earlier (Lines 2-3), and update labels whenever we identify an active ancestors (Lines 15-20). Our Pareto Search algorithm for edge weight increase is presented in Algorithms 4-5. Since shortest-path distances increase in the case of weight increase, we make the following changes: (1) replacing checks for shorter paths in the updated graph with checks for equal length paths passing through $(a, b)$ in the old graph (Line 17), and (2) marking labels as affected instead of updating them immediately (Lines 25-27), similar to Algorithm 2. However, this may lead to the repair phase becoming a bottleneck. This is because here paths of interest no longer need to pass through $(a, b)$, and thus repairs w.r.t. different ancestors can no longer be combined as we did during the search phase. To mitigate this, we take the following steps:

(1) We use $L_v[i] + \Delta$ as an upper bound for the value of $L_v[i]$ in the updated graph, and use it to repair $L_v[i]$ immediately (Line 18). Here $\Delta$ describes the weight increase of $(a, b)$.
(2) Instead of collecting individual affected pairs $(v, i)$ that need to be repaired, we group them and collect pairs $(v, [min, max])$ containing affected intervals (Lines 25-27).

The first step can reduce the number of repair operations when the upper bound used for immediate repairs is tight. This is most likely when the increase is small. While the second step will actually increase the number of distance comparisons during the initialization phase of the repair, it improves locality of reference for these checks and reduces edge traversals.

*Example 5.14.* Consider the weight of an edge $(1, 9)$ in Figure 2 decreases from 4 to 1. We perform two searches from the two endpoints 1 and 9 w.r.t. ancestors $\{2, 4, 5, 10, 9\}$. In Figure 6(a), we show the search from the endpoint vertex 1 and tracking distances to 9, with updated distance label values for ancestors 2 and 5. For vertices $\{1, 8, 10, 12\}$, the Pareto-active and active intervals are both $[0, 4]$, covering the label indices $\{0, 2, 4\}$ of active ancestors $\{2, 5, 9\}$ as well as label indices $\{1, 3\}$ of ancestors $\{4, 10\}$ which are inactive for this search. At vertex 4 the Pareto-active interval is reduced to $[0, 1]$, which is then pruned to the active interval $[0, 0]$. This interval is processed first at vertex 11 (due to shorter distance) before processing the active interval $[0, 4]$ originating from vertex 10. Here we first reduce $[0, 4]$ to $[1, 4]$ as label index 0 has already been processed. Now if the weight of an edge $(1, 9)$ increases from 4 to 7, we again run two searches but track distances in the original graph instead. In Figure 6(b), we show the search starting from vertex 1 and tracking distances to the root 9, with updates of distance labels for ancestors 2 and 5. As new distance values are not immediately available, we compute upper bounds first in Figure 6(b), and update them once we have found all affected vertices, as shown in Figure 6(c). Pareto-active and active intervals are both $[0, 4]$ for vertices $\{1, 8, 10, 12\}$, while vertices 4 and 11 are not affected.

## 6 THEORETICAL ANALYSIS

In the following we shall denote by $P_G^w(v, u)$ the set of shortest paths between $v$ and $u$ in $G[\text{Desc}(w)]$. In our notation $G' = G \oplus \Delta G$ describing the updated graph after applying $\Delta G$ on $G$. We use $d_{\max}$ for the maximum degree of vertices in $G$, and $h$ for the maximum number of ancestor vertices $\tau(v)$ in $G$ w.r.t. our tree hierarchy.

**Label Search Algorithm.** We briefly provide the following lemmas for showing correctness of label search algorithm.

LEMMA 6.1. *Denote by $L_\Delta^-$ the set of vertex pairs $(v, r)$ for which a pair $(d, v)$ enters $Q_r$ in Algorithm 1. Then $(v, r) \in L_\Delta^-$ iff $P_{G'}^r(v, r)$ contains strictly shorter paths than $P_G^r(v, r)$.*

LEMMA 6.2. *Denote by $L_\Delta^+$ the set of vertex pairs $(v, r)$ for which a pair $(d, v)$ enters $Q_r$ in Algorithm 2. Then $(v, r) \in L_\Delta^+$ iff $P_{G'}^r(v, r)$ differs from $P_G^r(v, r)$.*

Following these lemmas, one can show the following.

THEOREM 6.3. *Algorithms 1 and 2 operate in $O(|L_\Delta^-| \cdot d_{\max} \cdot \log |V|)$ and $O(|L_\Delta^+| \cdot d_{\max} \cdot \log |V|)$, respectively.*

**Pareto Search Algorithm.** The preceding lemmas can be used to show correctness of Pareto search algorithm.

LEMMA 6.4. *Denote by $V_\Delta^-$ the set of vertices $v$ for which line 22 is reached in Algorithm 3. Then $v \in V_\Delta^-$ iff $P_{G'}^r(v, r)$ contains strictly shorter paths than $P_G^r(v, r)$ for some ancestor $r$.*

LEMMA 6.5. *Denote by $V_\Delta^+$ the set of vertices $v$ for which line 23 is reached in Algorithm 4. Then $v \in V_\Delta^+$ iff $P_{G'}^r(v, r)$ differs from $P_G^r(v, r)$ for some ancestor $r$.*

As before, these lemmas can be used to analyze complexity.

THEOREM 6.6. *Algorithms 3 and 4 operate in $O(|V_\Delta^-| \cdot h + |L_\Delta^-| \cdot d_{\max} \cdot \log |V|)$ and $O(|V_\Delta^+| \cdot h + |L_\Delta^+| \cdot d_{\max} \cdot \log |V|)$, respectively..*

Despite the theoretical upper bounds being worse for Pareto search algorithm, it performs faster in practice as the factors $h$ and $L_\Delta$ tend to be over-estimates in Theorem 6.6.

Table 2: Summary of datasets.

| Network | Region | $|V|$ | $|E|$ | Memory |
|---|---|---|---|---|
| NY | New York City | 264,346 | 733,846 | 17 MB |
| BAY | San Francisco | 321,270 | 800,172 | 18 MB |
| COL | Colorado | 435,666 | 1,057,066 | 24 MB |
| FLA | Florida | 1,070,376 | 2,712,798 | 62 MB |
| CAL | California | 1,890,815 | 4,657,742 | 107 MB |
| E | Eastern USA | 3,598,623 | 8,778,114 | 201 MB |
| W | Western USA | 6,262,104 | 15,248,146 | 349 MB |
| CTR | Central USA | 14,081,816 | 34,292,496 | 785 MB |
| USA | United States | 23,947,347 | 58,333,344 | 1.30 GB |
| EUR | Western Europe | 18,010,173 | 42,560,279 | 974 MB |

## 7 EXPERIMENTS

We use STL-L$^-$ and STL-L$^+$ to denote our label search algorithms for edge weight decrease and increase, respectively. Similarly, STL-P$^-$ and STL-P$^+$ denote our Pareto search algorithms for edge weight decrease and increase, respectively.

**Hardware and code.** All the experiments are performed on a Linux server Intel Xeon W-2175 with 2.50GHz CPU, 28 cores, and 512GB of main memory. All the algorithms were implemented in C++20 and compiled using g++ 9.4.0 with the -O3 option. Our implementation is available at https://github.com/mufarhan/stable_tree_labelling.

**Datasets.** We use 10 undirected real road networks, nine of them are from the US and publicly available at the webpage of the 9th DIMACS Implementation Challenge [10] and one is from Western Europe managed by PTV AG [3]. Table 2 summarises these datasets where the largest dataset is the whole road network in the USA.

**State-of-the-art methods.** We compare our algorithms with the following state-of-the-art methods: 1) A dynamic algorithm called *Incremental Hierarchical 2-Hop Labelling (IncH2H)* [32], 2) A dynamic algorithm called *Dynamic Tree Decomposition-based Hub Labelling (DTDHL)* [30], and 3) A static algorithm called *Hierarchical Cut 2-Hop Labelling (HC2L)* [12]. For IncH2H and DTDHL, we use IncH2H$^-$, IncH2H$^+$ and DTDHL$^-$, DTDHL$^+$ to denote their algorithms for edge weight decrease and increase, respectively. We do not consider search-based methods [14, 22, 27]. Although search-based method may have smaller maintenance cost, their query performance is usually orders of magnitude slower than the labelling-sed methods considered in this paper.

The code for IncH2H, DTDHL and HC2L was kindly provided by their authors and implemented in C++. We use the same parameter settings as suggested by the authors of these methods, unless otherwise stated. We select the balance partition threshold $\beta = 0.2$ to construct stable tree hierarchies. When a method fails to produce results due to memory error, we denote it as "–".

**Test input generation.** To evaluate update time, for each network, we randomly sampled 10 batches, each containing 1,000 updates. For each update $(a, b, \phi)$ of batch $t$, we first increase its weight to $2.0 \times \phi$ to test the performance of weight increase case and then decrease (restored) its weight to the original (i.e., to $\phi$) to test the performance of weight decrease case. We also evaluate the update time with varying weights, specifically, using 9 randomly sampled batches, we first increase weights of updates $(a, b, \phi)$ of batch $t$ to $(t+1) * \phi$ and then restore their weights to the original i.e., $\phi$ to test the performance of weight increase and decrease case, respectively.

To evaluate query time, we randomly sampled 1,000,000 pairs of vertices in each road network. Following [21, 23], we also sampled sets of pairs containing short, medium and long range query pairs. Specifically, for each road network, we generate 10 sets of pairs $Q_1, Q_2, \ldots, Q_{10}$ as follows: we set $l_{min}$ to be 1000 meters, and set $l_{max}$ to be the maximum distance of any pair of vertices in the network. Let $x = (\frac{l_{max}}{l_{min}})^{1/10}$. For each $1 \leq i \leq 10$, we sample 10,000 query pairs to form each set $Q_i$, in which the distance of the source and target vertices for each query falls in the range $(l_{min} \cdot x^{i-1}, l_{min} \cdot x^i]$. For each algorithm, we report the average query processing time. Note that we shall refer sets of pairs $Q_1-Q_4$, $Q_5-Q_7$ and $Q_8-Q_{10}$ as short, medium and long range query sets, respectively. Finally, we compare the memory size of labelling produced by the state-of-the-art methods with our method STL.

### 7.1 Performance Comparison

We compare the performance of STL against the state-of-the-art methods in terms of update time, query time, and labelling size.

*7.1.1 Update Time.* We report the average update time over 10 batches in Table 3.

**Weight decrease.** Table 3 shows that our algorithm STL-P$^-$ is considerably faster than IncH2H$^-$ on all datasets, by up to an order of magnitude, and orders of magnitude faster compared with DTDHL$^-$. Particularly, STL-P$^-$ significantly outperforms IncH2H$^-$ on large datasets. Our method STL-L$^-$ is comparable with IncH2H$^-$, and an order of magnitude faster than DTDHL$^-$.

**Weight increase.** For the case of weight increase, Table 3 shows essentially the same trends between our algorithm STL-P$^+$, IncH2H$^+$ and DTDHL$^+$ as for the respective weight decrease algorithms, though STL-L$^+$ is slower than IncH2H$^+$. Note that all algorithms are slower than their counterparts for the weight decrease case. This is because the weight decrease case shortens paths, and thus new distance values are known that allows to immediately update labels. In the weight decrease case, computing new distance values requires additional computations.

*7.1.2 Query Time.* In Table 5, we first report the average query time over 1 million random distance queries for all datasets. We confirm that STL is the fastest on all datasets amongst the dynamic approaches, and only marginally slower than HC2L. Specifically, STL is 1.5-3 times faster compared with IncH2H and DTDHL. The main reason for this is that STL labels are significantly smaller, and fewer distance entries need to be processed to answer queries, at least for distant vertex pairs where only high level cuts are utilized. Additionally, identifying which label entries to compare is simpler, and label entries used are always consecutive in memory.

**Querying with varying distance.** In Figure 9, we report results for short, medium and long range query pairs on three large datasets CTR, USA and EUR to test the performance of STL against IncH2H and HC2L. The results for other datasets are similar. STL significantly outperforms IncH2H for long range query sets. Long range queries encounter significantly a small number of common ancestors because their lowest common ancestors generally lie at higher levels of hierarchy. For short range query sets, STL is slower or comparable to IncH2H. This is because the lowest common ancestors for short range queries are more likely to be at lower levels of the hierarchy causing a larger number of common ancestors to be explored in the labels. STL is slower than HC2L for short and medium range query sets because it only considers vertices in the lowest common ancestor node to answer queries.

Table 3: Comparison of update times between our methods and state-of-the-art methods.

| Network | Update Time - Decrease [ms] | | | | Update Time - Increase [ms] | | | |
|---|---|---|---|---|---|---|---|---|
| | STL-P$^-$ | STL-L$^-$ | IncH2H$^-$ | DTDHL$^-$ | STL-P$^+$ | STL-L$^+$ | IncH2H$^+$ | DTDHL$^+$ |
| NY | 0.845 | 1.978 | 2.006 | 11.40 | 1.712 | 3.561 | 2.900 | 13.87 |
| BAY | 0.917 | 1.788 | 1.769 | 8.899 | 1.695 | 3.233 | 2.498 | 14.53 |
| COL | 1.898 | 3.882 | 3.306 | 12.74 | 3.456 | 6.977 | 4.613 | 34.35 |
| FLA | 2.303 | 5.209 | 3.585 | 32.45 | 4.109 | 9.554 | 4.981 | 34.22 |
| CAL | 4.975 | 16.67 | 13.89 | 99.24 | 10.11 | 31.04 | 20.20 | 106.4 |
| E | 7.996 | 39.21 | 29.33 | 261.5 | 17.48 | 73.76 | 43.57 | 273.1 |
| W | 12.26 | 52.71 | 47.76 | 604.9 | 25.14 | 100.2 | 68.99 | 1,292 |
| CTR | 27.23 | 164.4 | 213.1 | 2,329 | 54.03 | 314.5 | 309.7 | 5,347 |
| USA | 32.67 | 216.4 | 239.8 | – | 82.78 | 412.9 | 356.3 | – |
| EUR | 13.68 | 68.25 | 66.97 | – | 61.57 | 131.4 | 96.63 | – |

Table 4: Comparison of labelling sizes and construction times between our method and state-of-the-art methods.

| Network | Labelling Size | | | | Construction Time [s] | | | | # Label Entries | | Tree Height | |
|---|---|---|---|---|---|---|---|---|---|---|---|---|
| | STL | HC2L | IncH2H | DTDHL | STL | HC2L | IncH2H | DTDHL | STL | IncH2H | STL | IncH2H |
| NY | 129 MB | 172 MB | 850 MB | 391 MB | 2 | 3 | 4 | 9 | 30 M | 99 M | 283 | 717 |
| BAY | 104 MB | 134 MB | 814 MB | 377 MB | 2 | 3 | 3 | 5 | 23 M | 93 M | 245 | 411 |
| COL | 175 MB | 238 MB | 1.37 GB | 587 MB | 4 | 6 | 5 | 7 | 40 M | 166 M | 386 | 556 |
| FLA | 423 MB | 561 MB | 2.43 GB | 1.30 GB | 11 | 16 | 11 | 17 | 97 M | 282 M | 276 | 496 |
| CAL | 1.03 GB | 1.48 GB | 8.21 GB | 3.91 GB | 28 | 44 | 30 | 48 | 251 M | 1.0 B | 481 | 722 |
| E | 2.92 GB | 4.22 GB | 20.7 GB | 9.68 GB | 75 | 129 | 74 | 111 | 735 M | 2.6 B | 560 | 1300 |
| W | 4.82 GB | 7.01 GB | 36.3 GB | 20.6 GB | 120 | 249 | 126 | 194 | 1.2 B | 4.5 B | 645 | 1115 |
| CTR | 19.7 GB | 30.2 GB | 178 GB | 80.3 GB | 540 | 1,140 | 858 | 766 | 5.0 B | 23 B | 1066 | 2522 |
| USA | 35.6 GB | 53.6 GB | 308 GB | – | 852 | 1,721 | 1,081 | – | 9.2 B | 40 B | 1181 | 2541 |
| EUR | 36.4 GB | 51.2 GB | 322 GB | – | 1,236 | 2,354 | 1,254 | – | 9.5 B | 42 B | 1429 | 3845 |

Table 5: Comparison of query times between our methods and state-of-the-art methods.

| Network | Query Time [$\mu$s] | | | |
|---|---|---|---|---|
| | STL | HC2L | IncH2H | DTDHL |
| NY | 0.287 | 0.264 | 0.913 | 0.852 |
| BAY | 0.299 | 0.258 | 0.841 | 0.785 |
| COL | 0.349 | 0.318 | 1.018 | 0.988 |
| FLA | 0.396 | 0.349 | 1.019 | 0.958 |
| CAL | 0.490 | 0.484 | 1.333 | 1.380 |
| E | 0.630 | 0.550 | 1.683 | 1.585 |
| W | 0.664 | 0.601 | 1.702 | 1.819 |
| CTR | 0.812 | 0.702 | 2.483 | 2.658 |
| USA | 0.834 | 0.734 | 3.428 | – |
| EUR | 1.185 | 0. 879 | 3.888 | – |

*7.1.3 Labelling Size.* Table 4 shows that the labelling sizes produced by STL is significantly smaller than the state-of-the-art methods. On the largest three datasets, the labelling size of STL is about 9 times smaller than IncH2H. This is because STL produces minimal cuts which are very small in practice, and thus the distance labels store a smaller number of label entries. We note that the difference in the number of label entries is not as large as the difference in labelling size (at most factor 4). The additional increase in labelling size for IncH2H is due to auxiliary data tracked to facilitate efficient updates. DTDHL uses the same tree hierarchy as IncH2H but tracks far less additional data, resulting in smaller labelling size. Compared to HC2L for which efficient maintenance is hard, STL uses less space, due to a reduction in cut size caused by the absence of shortcuts, though partially negated by the absence of tail-pruning.

## 7.2 Performance Analysis

Figure 8 shows how the average update time of our algorithms under edge weight decrease and increase behaves. We can see that the update time for our algorithm STL-P$^-$ and state-of-the-art algorithms IncH2H$^+$ and IncH2H$^-$ is independent of how much weights decrease or increase, while the update time for STL-P$^+$ increases as weights increase more. The variability of STL-P$^+$ performance can be traced back to line 18 of Algorithm 4: as the weight increase factor grows, the upper bound computed here will be tight less frequently, and more time is spent in function Repair. Except for COL and FLA, our algorithms STL-P$^+$ and STL-P$^-$ outperform IncH2H$^+$ and IncH2H$^-$ under both decrease and increase case. This has several reasons – for one, the number of labels for our approach is smaller than for IncH2H, as shown in Table 4, leading to fewer affected labels. Second, storing only distances within subgraphs reduces the number of labels affected by changes in $G$ even further. Finally, while IncH2H takes steps to ensure strong theoretical bounds for its update algorithms, such as tracking the support of nodes, we found that the practical impact of these is often limited.

We test the scalability of our algorithms STL-P$^+$ and STL-P$^-$ on the largest 3 datasets, CTR, USA and EUR. Following the same setting as described in the test input generation, we randomly sample 8,000 updates for each dataset and process them in groups of sizes ranging from 500 to 8,000, i.e., $\{5, 10, 15, 20, 25, 30, 35, 40, 45, 50, 55, 60, 65, 70, 75, 80\} \times 10^2$. We first process each group by applying the weight increases, followed by weight decreases, and then compare the results with the time required by STL to construct the labelling for each of these datasets (Table 4). As shown in Figure 10, even with the largest group (8,000 updates), the update times of STL-P$^+$ and STL-P$^-$ remain lower than the time required to fully reconstruct STL. Overall, STL-P$^+$ and STL-P$^-$ can process a significantly large amount of dynamic updates on a road network, enabling fast query processing without needing to rebuild the labels from scratch.

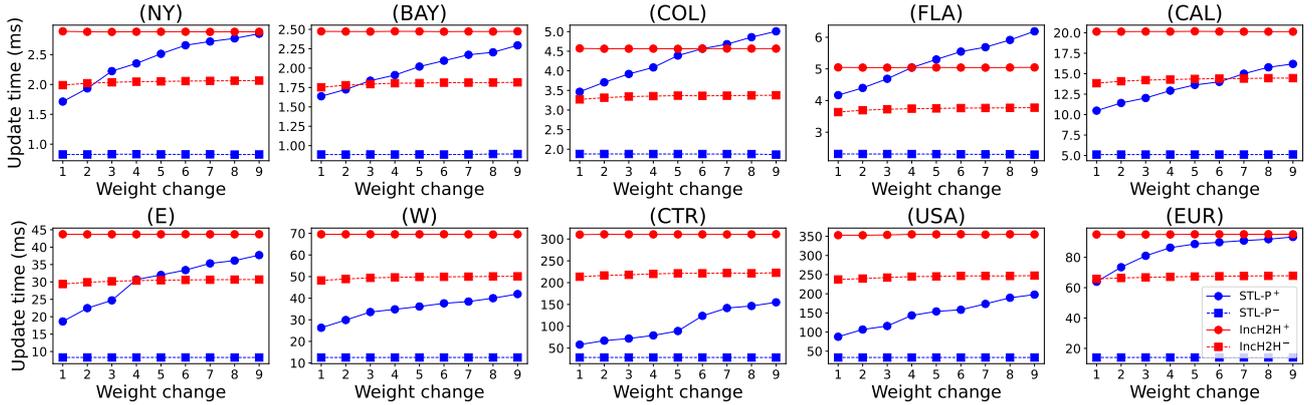

Figure 8: Update performance for both weight decrease and weight increase cases under varying edge weights.

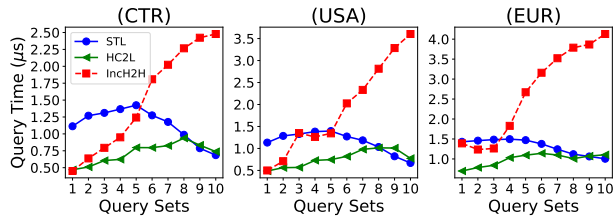

Figure 9: Query performance under varying distances.

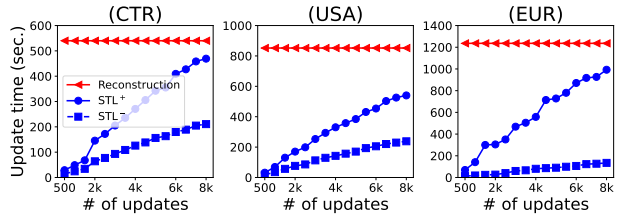

Figure 10: Update performance compared to reconstruction time for groups of updates ranging in size from 500 to 8,000.

## 8 EXTENSIONS

*Directed Road Networks.* Our algorithms can be easily extended to dynamic directed road networks. We may store distances from both directions in the label of each vertex $v \in V(G)$ when constructing STL. This can be achieved by performing searches in both directions during label construction. Then, our Label Search and Pareto Search algorithms can maintain STL using two Dijkstra's searches, namely forward and backward search. Specifically, Label Search algorithms perform such searches w.r.t. each ancestor and Pareto Search algorithms conduct them w.r.t. each edge whose weight is increased or decreased to maintain STL for the directed version.

*Edge/Vertex Insertion/Deletion.* In practice, new roads are seldom built and old roads are rarely deconstructed. Thus, the structure of road networks is considered to be intact in general. As a result, structural changes such as edge or vertex insertion and deletion in road networks are extremely infrequent. Prior work has addressed such changes [21, 30, 32]. Similarly, our algorithms can be adapted to handle these changes within the STL framework as follows. An edge deletion can be handled by increasing the weight of the deleted edge to $\infty$ and similarly a vertex deletion can be handled by increasing the weights of its adjacent edges to $\infty$. For edge insertions, we can identify the affected nodes in the stable tree hierarchy, re-partition their induced subgraphs, and fix the affected tree nodes at the lower levels. Afterward, we compute new labels for these tree nodes using the algorithms in [12].

## 9 CONCLUSION

In this paper, we tackled the challenge of maintaining distance labeling to efficiently answer shortest-path queries on dynamic road networks. We introduced the concepts of a *stable tree hierarchy* and *stable tree labeling (STL)*, which serve as the foundation for designing efficient dynamic algorithms. Then, we developed two novel algorithms: the Label Search and Pareto Search algorithms, from different perspectives. Label Search is ancestor-centric, focusing on efficiently updating labels related to ancestors, while Pareto Search is update-centric, optimizing updates to the labeling by eliminating duplicate search traversals. These algorithms can significantly reduce the search space involved in maintaining STL. Our experiments, conducted on 10 large real-world road networks, demonstrated that the proposed algorithms substantially outperform existing approaches in terms of both query processing and update time, showcasing their practical effectiveness in dynamic settings.

A potential avenue for future work is to adapt STL to other dynamic graph structures, such as social or communication networks. These networks feature distinct characteristics, such as high clustering coefficients or fluctuating connectivity, which present new challenges and opportunities for optimizing label construction and maintenance. This exploration could lead to enhancements in handling dynamic graphs across various domains.